\documentclass[aps,pre,reprint,groupedaddress]{revtex4-1}
\usepackage{amsmath}

\usepackage{graphicx}
\usepackage{epstopdf}

\begin{document}


\title{Efficient Bayesian estimation of permutation entropy with Dirichlet priors}


\author{Douglas J. Little, J. P. Toomey and Deb M. Kane}
\affiliation{MQ Photonics Research Centre, Department of Physics and Astronomy, Macquarie University, North Ryde, NSW 2109, Australia}


\date{\today}

\begin{abstract}
	
Estimation of permutation entropy (PE) using Bayesian statistical methods is presented for systems where the ordinal pattern sampling follows an independent, multinomial distribution. It is demonstrated that the PE posterior distribution is closely approximated by a standard Beta distribution, whose hyperparameters can be estimated directly from moments computed analytically from observed ordinal pattern counts. Equivalence with expressions derived previously using frequentist methods is also demonstrated. Because Bayesian estimation of PE naturally incorporates uncertainty and prior information, the orthodox requirement that $N \gg D!$ is effectively circumvented, allowing PE to be estimated even for very short time series. Self-similarity tests on PE posterior distributions computed for a  semiconductor laser with optical feedback (SLWOF) system show its PE to vary periodically over time.

\end{abstract}


\maketitle


\section{Introduction}

Permutation entropy is an information-theoretic quantity that has been applied to a diverse array of complex systems including chaotic laser systems \cite{Soriano11, Toomey14}, speech patterns \cite{Bandt02}, heart (ECG) and brain (EEG) activity \cite{Frank06, Parlitz12, Olofsen08}, stock market indices \cite{Zunino09}, and more \cite{Riedl13}. Permutation entropy is derived, first by breaking the time series into small parts (words), then assigning each word a symbol, based on where the highest-valued, lowest-valued etc. elements are located within the word. The permutation entropy is the Shannon entropy of this encoded symbol string.

This statistic, first defined by Bandt and Pompe \cite{Bandt02}, is designed to quantify the predictability of dynamical system states. In it's simplest incarnation (word length of 2), it quantifies the probability a time series is increasing or decreasing at any given point in time. Longer word lengths are able to capture more complex dynamics, e.g. a word length of 3 can distinguish between monotonic and convex/concave patterns. In addition to the word length (the embedding dimension, $D$), there is also a native dependence on the sampling frequency, which can be varied via the embedding delay, $\tau$.

The attraction of using permutation entropy versus other complexity measures is the sensitivity with which it can detect shifts in underlying system behavior, the simplicity and speed with which it can be computed, its near-universal applicability to ordered data sets, its robustness toward detector nonlinearities and noise, and the ease with which it can be interpreted, given its similarity with the well-known Shannon entropy. 

While PE is often viewed as a property of data sets, it is reasonable to extend this association to the physical systems that produce the data. Just as the proportion of heads in a sequence of coin flips is connected to the physical properties of the coin, so too can the PE of a data set be viewed as a physical reflection of the generating system. Indeed, the principal application of PE has been to discriminate physical states by observing statistically-significant changes in the observational frequency of output patterns/symbols.

A key aspect of this process is determining what constitutes statistical significance. The fairness of a coin cannot be exactly determined from a finite number of tosses. Similarly PE of a physical system cannot be exactly determined from a finite data set. In this paradigm, it can only be estimated with a finite confidence interval. Despite its wide application of PE, methods for evaluating of these confidence intervals are somewhat sparse \cite{Little16, Little17, Traversaro18, Pose21}. In large part, this is because experimenters largely dealt with ensembles of physical systems, and so the PE confidence interval could be made negligible compared the ensemble standard deviation by enforcing the condition that $N \gg D!$. Reliance on this blanket rule however overlooks useful information that can potentially be extracted from smaller data sets, and inhibits self-similarity testing between individual systems. 

Bayesian statistical methods have gained significant popularity in recent decades and have become ubiquitous in contemporary fields such as machine learning \cite{Bishop06}. In Bayesian statistics, parameters are represented by probability distributions that represent degrees of belief. These distributions evolve with the acquisition of new data, reflecting the gain in information. Determining permutation entropy variances is achieved naturally by expressing the permutation entropy as a probability distribution, and then evaluating the associated moments. This can be done with extremely limited data sets (10s or 100s of data points) due to the incorporation of prior distributions within the Bayesian framework.   

In this paper, the use of Bayesian methods to estimate permutation entropy will be explored for independent, multinomial distributions. Independence in this context means that symbol probabilities are uncorrelated with previous observations. In section \ref{sec:theory}, the concept of permutation entropy is introduced along with Bayes' rule, and the important role of the Dirichlet distribution in the case of multinomial sampling. From this, it is shown how computation of \textit{moments} can expedite computations by adapting the methods of Wolpert and Wolf \cite{Wolpert95}, to permutation entropy distributions. Section \ref{sec:PW} demonstrates this Bayesian approach to be a generalization of results generated using frequentist methods. In section \ref{sec:RD}, these methods are tested via numerical simulation, and then applied to time series generated by a semiconductor laser system subject to optical feedback; a system known to produce complex behavior \cite{Toomey14}. Conclusions and final outlooks are provided in section \ref{sec:conclusion}.

\section{Theory}

\label{sec:theory}

\subsection{Permutation Entropy}

Permutation entropy (PE) is a statistic designed to characterize the dynamics of a system. The first step to computing the PE is to partition time series elements into vectors (words). Given a time series $\{x_t\}, t = 1,\ldots,L$, it can be partitioned into a series of vectors comprising of $D$ consecutive elements such that;
\begin{equation}
\mathbf{x}_s = \{x_{(s-1)D+1},\ldots,x_{sD}\},
\end{equation}
where $D$ is a free parameter called the embedding dimension (the word length). Here, the time series has been partitioned so that each time series element belongs to (at most) one vector. If $D = 4$ is chosen for example, $\mathbf{x}_1 = \{x_1,\ldots,x_4\}$, $\mathbf{x}_2 = \{x_5,\ldots,x_8\}$, and so forth. The total number of vectors, $N$ (i.e. the maximum $s$), is $\lfloor L/D\rfloor$. This partitioning is designed to avoid introducing additional correlations between observed dynamical states \cite{Little17}.

The next step is to associate a dynamical state with each vector, which are assigned based on the rank sequence. A vector has the rank sequence $[X_1,\ldots,X_D]$ if it satisfies the condition;
\begin{equation}
x_{X_1} < \ldots < x_{X_D}
\end{equation}
where subscripts indicate the position of the time series element within the vector (from 1 to $D$). In this paper, random numbers are used to break equalities. As an example, the vector $\{1, 10, 100, 99\}$ would be assigned the rank sequence $[1, 2, 4, 3]$ since $x_1 < x_2 < x_4 < x_3$. There are $D!$ possible rank permutations and thus $D!$ different dynamical states, identified with the symbols $\pi_i$. 

The permutation entropy (PE) is defined as;
\begin{equation}
\label{eq:PE}
H = -\frac{1}{\log(D!)}\sum_{i=1}^{D!}P_i\log(P_i),
\end{equation}
where $\mathbf{P} = \{P_1,\ldots,P_{D!}\}$ is the discrete (marginal) probability distribution that an ordinal pattern vector will be assigned the symbol $\pi_i$. The normalization term is included to ensure $H$ falls on the bounded interval $[0, 1]$, which facilitates comparison of $H$ for different $D$. Note that some authors refer to the PE as the entropy of $\mathbf{P}$ absent the normalization term. In this paper, PE always refers to the normalized quantity in Eq. \ref{eq:PE}.

Exact calculation of PE without \textit{a priori} knowledge of $\mathbf{P}$ requires an infinite number of observations, so an estimator of the form 
\begin{equation}
\label{eq:PEest}
H_\mathrm{est} = -\frac{1}{\log(D!)}\sum_{i=1}^{D!}\frac{p_i}{N}\log\left(\frac{p_i}{N}\right),
\end{equation}
is generally used, where $p_i$ is the count of observed dynamical states and $N = \sum p_i$ is the total number of observations. It is readily seen that Eq. \ref{eq:PEest} approaches Eq. \ref{eq:PE} in the limit of large $N$.

If symbol observations are modeled as samples randomly drawn from $\mathbf{P}$, $p_i$ will have a finite variance, and thus $H_{\mathrm{est}}$ will be drawn from a probability distribution $p(H)$. One approach to determine $p(H)$ is through repeated acquisitions of data, however this may not be feasible due to data/memory requirements, or if $\mathbf{P}$ is non-stationary.

The alternate, Bayesian approach is to compute $p(H)$ for a given set of observations, by assigning a probability for each possible $\mathbf{P}$ based on the observed symbols. Central to the Bayesian approach is Bayes' rule, which can be summarized in this context as
\begin{equation}
\label{eq:Bayes}
p(\mathbf{P}|O) = \frac{p(O|\mathbf{P})p(\mathbf{P})}{p(O)},
\end{equation}
where $O$ is some set of observed dynamic states. The basic idea is to take an existing \textit{prior} distribution over $\mathbf{P}$ and use Bayes' rule to obtain an updated probability distribution, $p(\mathbf{P}|O)$ (the posterior distribution), that incorporates the information contained in $O$. The likelihood function, $p(O|\mathbf{P})$, essentially defines the random process used to model the sequence of observations contained in $O$. The change of variable defined in Eq. \ref{eq:PE} can be used to retrieve $p(H|O)$ from $p(\mathbf{P}|O)$. 

\subsection{Applying Bayes' rule to iid systems}

An independent, identically-distributed (iid) system is defined here as a process where $\mathbf{P}$ is independent of previous observations. In this instance the likelihood function has a simple form
\begin{equation}
\label{eq:likelihood}
p(O|\mathbf{P}) = \prod_{i=1}^{D!}P_i^{p_i}.
\end{equation}
It is well-established practice in this case to use a so-called \textit{conjugate} prior in the form of a Dirichlet distribution \cite{Raiffa61, Bishop06},
\begin{equation}
\label{eq:Dirichlet}
p(\mathbf{P}) = \frac{\Gamma(\alpha_0)}{\Gamma(\alpha_1)\ldots\Gamma(\alpha_{D!})}\prod_{i=1}^{D!}P_i^{\alpha_i-1}.
\end{equation}
where $\alpha_i$ are hyperparameters of the Dirichlet distribution, $\alpha_0 = \sum_i^{D!}\alpha_i$ and $\Gamma$ represents the Gamma function. This expedites the evaluation of Eq. \ref{eq:Bayes} by ensuring the prior and posterior distributions have the same functional form. Using Eq. \ref{eq:likelihood} and \ref{eq:Dirichlet} in conjunction with Eq. \ref{eq:Bayes}, the posterior distribution is
\begin{equation}
\label{eq:posterior}
p(\mathbf{P}|O) = \frac{\Gamma(\alpha_0 + N)}{\Gamma(\alpha_1+p_1)\ldots\Gamma(\alpha_{D!}+p_{D!})}\prod_{i=1}^{D!}P_i^{\alpha_i+p_i-1}
\end{equation}
where the normalization terms can be determined by inspection. From Eq. \ref{eq:posterior}, the hyperparameters of the posterior distribution are found by adding the observed symbol counts $p_i$, to the corresponding hyperparameters of $p(\mathbf{P})$. Thus $\alpha_i$ are commonly interpreted as equivalent symbol counts ``worth" of prior information.

Computing $p(H|O)$ directly from $p(\mathbf{P|O})$ is computationally challenging due to the nonlinear change-of-variable and the high-dimensionality of $\mathbf{P}$. A more efficient approach explored here is to reconstruct $p(H|O)$ from its computed moments. Because $H$ lies on a closed, finite interval, the moments of $p(H|O)$ uniquely determine its functional form \cite{Hausdorff21}.

\subsection{Calculating moments}

Prior distributions will be considered initially here for the sake of brevity. As will be seen, the extension to posterior distributions is straightforward. The $n$th moment of $p(H)$ is defined as
\begin{equation}
E[H^n] = \int_{0}^{1}H^np(H)dH.
\end{equation}
This is equivalent to (via change of variable)
\begin{equation}
\label{eq:EH}
E[H^n] = \frac{(-1)^n}{\log(D!)^n}\int_{S}\left(\sum_{i=1}^{D!}P_i\log(P_i)\right)^np(\mathbf{P})d\mathbf{P},
\end{equation}
where the integral is taken over the simplex, $S$, defined by $\sum P_i=1$. Terms arising from the summation raised to the power of $n$ can be grouped according to the multinomial theorem
\begin{equation}
\left(\sum_{i=1}^{D!}P_i\log(P_i)\right)^n = \sum_{|\mathbf{k}|=n}{n\choose \mathbf{k}}\mathbf{P}^\mathbf{k}\log(\mathbf{P})^\mathbf{k},
\end{equation}
where $\mathbf{k}$ is a $D!$-dimensional multi-index, which has the property $\mathbf{x^k} = x^{k_1}\ldots x^{k_{D!}}$, and ${n\choose \mathbf{k}}$ represents multinomial coefficients. Eq. \ref{eq:EH} can be recast by exchanging the order of the integral and the summation over $|\mathbf{k}| = n$ as
\begin{equation}
\label{eq:moment}
E[H^n] = \frac{(-1)^n}{\log(D!)^n}\sum_{|\mathbf{k}|=n}\int_{S}{n\choose \mathbf{k}}\mathbf{P}^\mathbf{k}\log(\mathbf{P})^\mathbf{k}p(\mathbf{P})d\mathbf{P}.
\end{equation}
Eq. \ref{eq:moment} can be reduced to integrals of the form
\begin{equation}
\label{eq:multi}
\frac{\Gamma(\alpha_0)}{\Gamma(\alpha_1)\ldots\Gamma(\alpha_{D!})}\int_S\mathbf{P}^\mathbf{k}\log(\mathbf{P})^\mathbf{k}\prod_{j=1}^{D!}P_j^{\alpha_j-1}d\mathbf{P},
\end{equation}
by substituting the explicit form of $p(\mathbf{P})$ (Eq. \ref{eq:Dirichlet}). These integrals can be solved by first expressing the integrand as a derivative
\begin{equation}
\label{eq:multi2}
\frac{\Gamma(\alpha_0)}{\Gamma(\alpha_1)\ldots\Gamma(\alpha_{D!})}\int_S\frac{\partial^n}{\partial\boldsymbol{\alpha}^\mathbf{k}}\prod_{j=1}^{D!}P_j^{\alpha_j-1+k_j}d\mathbf{P},
\end{equation}
then using the normalization condition for the Dirichlet distribution to express Eq. \ref{eq:multi2} as
\begin{equation}
\frac{\Gamma(\alpha_0)}{\Gamma(\alpha_1)\ldots\Gamma(\alpha_{D!})}\frac{\partial^n}{\partial\boldsymbol{\alpha}^\mathbf{k}}\frac{\Gamma(\alpha_1+k_1)\ldots\Gamma(\alpha_{D!}+k_{D!})}{\Gamma(\alpha_0+n)}.
\end{equation}
The $n$th moment of $H$ can then be expressed as
\begin{equation}
\label{eq:Betas}
E[H^n] = \frac{(-1)^n}{\log(D!)^n}\sum_{|\mathbf{k}|=n}{n\choose \mathbf{k}}\frac{1}{\mathbf{B}(\boldsymbol{\alpha})}\frac{\partial^n}{\partial\boldsymbol{\alpha}^\mathbf{k}}\mathbf{B}(\boldsymbol{\alpha}+\mathbf{k}),
\end{equation}
where $\mathbf{B}$ is the multivariate Beta function.

Calculating the first moment (i.e. the mean), $n = |\mathbf{k}| = 1$, hence the summation is over multi-indices where one component is 1 and all the other components are zero. Eq. \ref{eq:Betas} therefore reduces to a sum of first derivatives;
\begin{equation}
E[H] = -\frac{1}{\log(D!)}\sum_{i=1}^{D!}\frac{\Gamma(\alpha_0)}{\Gamma(\alpha_i)}\frac{\partial}{\partial\alpha_i}\frac{\Gamma(\alpha_i+1)}{\Gamma(\alpha_0+1)}.
\end{equation}
Resolving the derivatives yields
\begin{equation}
\label{eq:FM}
E[H] = -\frac{1}{\log(D!)}\sum_{i=1}^{D!}\frac{\alpha_i}{\alpha_0}(\psi(\alpha_i+1)-\psi(\alpha_0+1)),
\end{equation}
where $\psi$ is the digamma function. Thus the expectation of $H$ can be determined directly from the hyperparameters of the Dirichlet distribution $p(\mathbf{P})$, without the need for statistical sampling methods.

For the second moment, $n = |\mathbf{k}| = 2$. Here, the summation in Eq. \ref{eq:Betas} can be separated into two parts; the first containing multi-indices where one component is 2 and the others are zero, and the second containing multi-indices where two components are 1 and the rest zero. Separating the summation in this fashion yields the expression
\begin{equation}
\begin{split}
\label{eq:SMdiv}
E[H^2] = \frac{1}{\log(D!)^2}\bigg[\sum_{i=1}^{D!}\frac{\Gamma(\alpha_0)}{\Gamma(\alpha_i)}\frac{\partial^2}{\partial\alpha_i^2}\frac{\Gamma(\alpha_i+2)}{\Gamma(\alpha_0+2)}\\+ \sum_{i,j=1;i\neq j}^{D!}\frac{\Gamma(\alpha_0)}{\Gamma(\alpha_i)\Gamma(\alpha_j)}\frac{\partial^2}{\partial\alpha_i\alpha_j}\frac{\Gamma(\alpha_i+1)\Gamma(\alpha_j+1)}{\Gamma(\alpha_0+2)}\bigg].
\end{split}
\end{equation}
Note that the coefficient of the second term depends on whether this summation is over combinations of $i$ and $j$, or permutations. Summing over permutations is most natural from a computing perspective, so that is the convention chosen here. The coefficient of the second summation is therefore 1, not the multinomial index value of 2 as it would be if the summation was over combinations.

Evaluating the derivatives in Eq. \ref{eq:SMdiv} yields
\begin{widetext}
\begin{equation}
\begin{split}
\label{eq:SM}
E[H^2] = \frac{1}{\log(D!)^2}\bigg[\sum_{i=1}^{D!}\frac{\alpha_i(\alpha_i+1)}{\alpha_0(\alpha_0+1)}\bigg((\psi(\alpha_i+2)-\psi(\alpha_0+2))^2+\psi_1(\alpha_i+2)-\psi_1(\alpha_0+2)\bigg)\\ +\sum_{i,j=1;i\neq j}^{D!}\frac{\alpha_i\alpha_j}{\alpha_0(\alpha_0+1)}\bigg((\psi(\alpha_i+1)-\psi(\alpha_0+2))(\psi(\alpha_j+1)-\psi(\alpha_0+2))-\psi_1(\alpha_0+2)\bigg)\bigg],
\end{split}
\end{equation}
\end{widetext}
where $\psi_1$ is the trigamma function. The variance is determined from the first two moments as
\begin{equation}
\label{eq:Var}
\mathrm{Var}[H] = E[H^2]-E[H]^2.
\end{equation}

Equivalent expressions for the moments of the posterior distribution can be found by replacing the hyperparameters $\alpha_i$ with $\alpha_i + p_i$. This analysis can be extended to higher order moments through judicious separation of the summation in Eq. \ref{eq:Betas} and evaluation of the relevant derivative terms. Expressions for the third and fourth standardized moments (skewness and kurtosis) and their derivations can be found in the appendix and a script for performing these calculations is provided as additional material. Note that polygamma functions, and thus the moments of $p(H)$, can be efficiently evaluated on most modern numerical computing platforms. The next step is to construct $p(H)$ from computed moments, this will be discussed in section \ref{sec:RD}.

\section{Comparisons to previous work}

\label{sec:PW}

The mean and variance of $p(H)$ have previously be computed through frequentist statistics \cite{Little16, Little17}. Here, it will be demonstrated that these expressions are special cases of the more general Bayesian approach. First, the expectation of $H$, Eq. \ref{eq:FM}, will be approximated in the limit of large $N$. It is noted first that large $N$ implies large posterior hyperparameters. As the digamma function can be approximated as
\begin{equation}
\label{eq:digamma}
\psi(x) \approx \log(x)-\frac{1}{2x},
\end{equation}
for large $x$, this can be combined with the linear approximation for the derivative of a logarithm
\begin{equation}
\log(x+1)-\log(x) \approx \frac{2}{2x+1},
\end{equation}
to get the following approximation to Eq. \ref{eq:FM}
\begin{equation}
\begin{split}
E[H] \approx -\frac{1}{\log(D!)}\sum_{i=1}^{D!}\frac{\alpha_i}{\alpha_0}\bigg(\log\left(\frac{\alpha_i}{\alpha_0}\right) +\\\frac{2}{2\alpha_i+1}-\frac{1}{2\alpha_i+2}-\frac{2}{2\alpha_0+1}+\frac{1}{2\alpha_0+2}\bigg).
\end{split}
\end{equation}
In the case of a uniform prior, $\alpha_i = p_i$ and $\alpha_0 = N$. This expression is equivalent to
\begin{equation}
E[H] \approx H_\mathrm{est}-\frac{1}{2N\log(D!)}\sum_{i=1}^{D!}1-\frac{p_i}{N},
\end{equation}
which is equivalent to the expression in Ref. \cite{Little17} for the expectation of $H$ at large $N$
\begin{equation}
\label{eq:prevFM}
E[H] = H_\mathrm{est}-\frac{D!-1}{2N\log(D!)}.
\end{equation}
This demonstrates that Eq. \ref{eq:FM} is a more general expression for $E[H]$, applicable to any $N$ and where the Dirichlet prior is not necessarily uniform.

The second moment (Eq. \ref{eq:SM}) can be related to the variance using Eq. \ref{eq:Var}. Using the expression for $E[H]$ in Eq. \ref{eq:FM}, applying the approximate recurrence relations
\begin{equation}
\begin{split}
\psi(x+1) &\approx \psi(x)+\frac{1}{x},\\
\psi_1(x+1) &\approx \psi(x)-\frac{1}{x^2},
\end{split}
\end{equation}
and grouping the $\psi(\alpha_i+1)-\psi(\alpha_0+1)$ terms 
\begin{widetext}
\begin{equation}
\begin{split}
\label{eq:Varapprox}
\mathrm{Var}[H] &= \frac{1}{\log(D!)^2}\sum_{i=1}^{D!}\frac{\alpha_i(\alpha_i+1)}{\alpha_0(\alpha_0+1)}\left(\Delta\psi_1(\alpha_i + 1)+\frac{2}{(\alpha_0+1)^2}+\left(1-\frac{\alpha_i(\alpha_0+1)}{\alpha_0(\alpha_i+1)}\right)\Delta\psi(\alpha_i + 1)^2-\frac{2}{(\alpha_i+1)(\alpha_0+1)}\right)\\
&+\frac{1}{\log(D!)^2}\sum_{i,j=1;i\neq j}^{D!}\frac{\alpha_i\alpha_j}{\alpha_0(\alpha_0+1)}\left(\frac{2}{(\alpha_0+1)^2}-\frac{1}{\alpha_0}\Delta\psi(\alpha_i + 1)\Delta\psi(\alpha_j + 1)-\psi_1(\alpha_0+1)\right),
\end{split}
\end{equation}
\end{widetext}
where the notation $\Delta\psi_m(x + M) = \psi_m(x + M)-\psi_m(\alpha_0+2)$ is adopted for brevity. Approximating the digamma and trigamma functions, using Eq. \ref{eq:digamma} and
\begin{equation}
\psi_1 \approx \frac{1}{x}+\frac{1}{2x^2} 
\end{equation}
respectively, using the relation $\sum\alpha_i = \alpha_0$, and omitting less significant terms, Eq. \ref{eq:Varapprox} can be approximated as;
\begin{equation}
\label{eq:Varapprox2}
\begin{split}
\mathrm{Var}[H] &= \frac{1}{(\alpha_0+1)\log(D!)^2}\bigg[\sum_{i=1}^{D!}\frac{\alpha_i}{\alpha_0}\left(1-\frac{\alpha_i}{\alpha_0}\right)\log\left(\frac{\alpha_i}{\alpha_0}\right)\\ &- \sum_{i,j=1;i\neq j}^{D!}\frac{\alpha_i\alpha_j}{\alpha_0^2}\log\left(\frac{\alpha_i}{\alpha_0}\right)\log\left(\frac{\alpha_j}{\alpha_0}\right)
\\ &+\frac{1}{4(\alpha_0+1)}\bigg(10+\sum_{i=1}^{D!}\frac{\alpha_i(\alpha_0+1)}{\alpha_0(\alpha_i+1)}-12\frac{\alpha_i}{\alpha_0}\bigg)\bigg].
\end{split}
\end{equation}
It is noted at this point that Eq. \ref{eq:Varapprox2} is dominated by the first two terms unless $\mathbf{P}$ is uniform, which implies that
\begin{equation}
\frac{\alpha_i(\alpha_0+1)}{\alpha_0(\alpha_i+1)} \approx 1
\end{equation}
when $N$ (equivalently, $\alpha_0$) is sufficiently large. Algebra then shows Eq. \ref{eq:Varapprox2} to be equivalent to Eq. 12 in \cite{Little17} (noting the surprising result that the 2nd and 4th terms in Eq. 12 \cite{Little17} combined are indeed equivalent to the final term in Eq. \ref{eq:Varapprox2}, independent of $\mathbf{P}$).

\section{Results and Discussion}

\label{sec:RD}

\subsection{Validating moment calculations}

The uniform Dirichlet distribution, characterized by the hyperparameter values $\alpha_i = 1$, arises when there is complete uncertainty about $\mathbf{P}$. That is, any value of $\mathbf{P}$ is just as likely as any other. It is the least informative of the Dirichlet distributions, and is therefore the default prior distribution when there is no \textit{a priori} knowledge of the system. To validate moments of this Dirichlet prior calculated analytically using Eqs. \ref{eq:FM}, \ref{eq:SM}, \ref{eq:TM} and \ref{eq:FoM}, they were compared against numerical simulations. For each simulation, $10^6$ samples were drawn from $p(\mathbf{P})$. $H$ was then determined for each sample, whereupon the moments of $p(H)$ were computed numerically from the resultant histogram. 

Comparisons between analytic and numerically calculated moments are shown in Table \ref{tab:Uniform_D3}. Agreement is well within 1\%, validating the accuracy of moments computing using the Bayesian approach outlined in the previous section. Discrepancies can mostly be attributed to sampling error. As $D$ increases, it can be seen that $\gamma$ and $\kappa$ tend toward 0 and 3 respectively, indicating that $p(H)$ tends toward a Gaussian distribution.
\begin{table}
	\caption{\label{tab:maintab}Analytically- and numerically-calculated standardized moments for $p(H)$ of the Dirichlet distribution with $\alpha_i = 1$.}
	\begin{tabular}{lccccc}
		&$D$& $\mu$& $\sigma^2$& $\gamma$& $\kappa$\\
		\hline
		Analytic&3&0.8091&0.009585&-0.9160&4.2219\\
		Numeric&3&0.8093&0.009569&-0.9120&4.2050\\
		Analytic&4&0.8735&0.001065&-0.5975&3.6928\\
		Numeric&4&0.8735&0.001068&-0.6014&3.7034\\
		Analytic&5&0.9126&0.000103&-0.2895&3.1665\\
		Numeric&5&0.9126&0.000103&-0.2921&3.1785\\
	\end{tabular}	
\end{table}

The accuracy of analytically-computed moments was found to improve for larger values of $\alpha_i$, hence the performance for $\alpha_i = 1$ can be considered a minimum baseline.

\subsection{Reconstructing $p(H)$ from computed moments}
\label{sec:Reconstructing}

\begin{figure}
	\includegraphics[width=8.6cm]{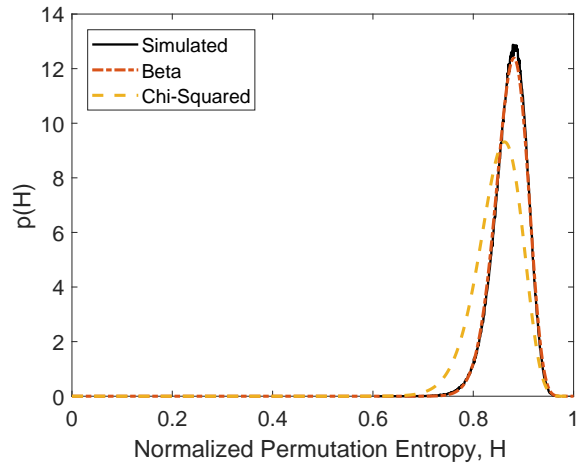}
	\caption{\label{fig:Uniform_D4} $p(H)$ for the Dirichlet distribution with $D = 4$ and $\alpha_i = 1$. Simulated data was acquired with $10^6$ samples of the Dirichlet distribution, with $H$ calculated for each. The Beta distribution fit was made using the analytically-calculated mean and variance of $p(H)$.}
\end{figure}

From analysis of computed moments, it was determined that standard Beta distributions provide the closest fit to $p(H)$. The Beta distribution has the form 
\begin{equation}
p(H) = \frac{\Gamma(\beta_1+\beta_2)}{\Gamma(\beta_1)\Gamma(\beta_2)}H^{\beta_1-1}(1-H)^{\beta_2-1},
\end{equation}
where $\beta_1, \beta_2$ are free (hyper-)parameters determined from the mean and variance via
\begin{equation}
\begin{split}
E[H] &= \frac{\beta_1}{\beta_1+\beta_2}\\
\mathrm{Var}[H] &= \frac{\beta_1\beta_2}{\left(\beta_1+\beta_2\right)^2\left(\beta_1+\beta_2+1\right)}.
\end{split}
\end{equation}
The closeness of the Beta distribution fit to the simulated data is demonstrated in Fig. \ref{fig:Uniform_D4}. The Beta distribution is also demonstrated to be an improvement at low $N$ over the $\chi^2$-distribution previously used to approximate $p(H)$ when $N \gg D!$ \cite{Little16}. 

Evaluating moments beyond the first two is useful for numerically validating the quality Beta distribution fit. Quantitatively, the skewness of the Beta distribution depicted in Fig. \ref{fig:Uniform_D4} was found to be -0.4354, slightly underestimating the skewness of the numerical data (Table \ref{tab:maintab}). This is reflected by the Beta distribution fit slightly underestimating the peak and overestimating the tails of the numerical data. 

\begin{figure}
	\includegraphics[width=8.6cm]{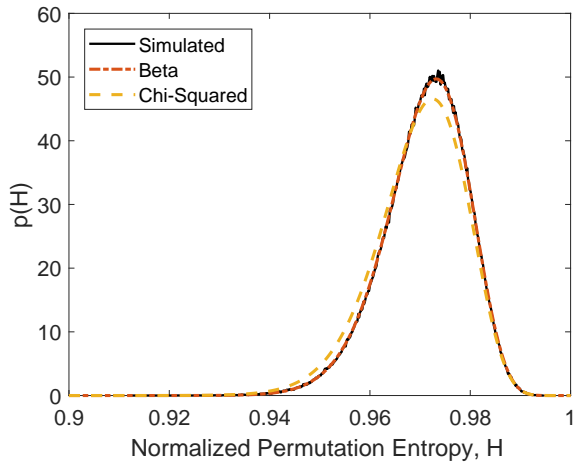}
	\caption{\label{fig:Symmetric_D4_5} $p(H)$ for the Dirichlet distribution with $D = 4$ and $\alpha_i = 5$. Simulated data was acquired with $10^6$ samples of the Dirichlet distribution, with $H$ calculated for each. The Beta distribution fit was made using the analytically-calculated mean and variance of $p(H)$.}
\end{figure}

The Dirichlet distribution where $\alpha_i = 5$ is shown in Fig. \ref{fig:Symmetric_D4_5} for comparison. Even with this small number of additional (equivalent) observations, the $\chi^2$-distribution is a much improved fit, consistent with it being intended for use at large $N$. The Beta distribution fit however remains the superior approximation for $p(H)$, and is visually indistinguishable from the numerically computed distribution in Fig. \ref{fig:Symmetric_D4_5}, with their skewnesses differing by just 0.05, compared to about 0.17 in the case where $\alpha_i = 1$.

In the general case, $\alpha_i$ don't have a fixed relationship. It is remarked that the Beta distribution remains a good fit for $p(H)$ in these instance even when $\alpha_i$ are highly asymmetric (i.e. a few large values interspersed among smaller values). In fact, the only notable trend was that the Beta distribution fit improves the larger $\alpha_0$ becomes. The only observed circumstances where the Beta fit was not particularly good was when $\alpha_i$ occupied values less than unity, which can arise for prior distributions where $\mathbf{P}$ is confined towards an edge or vertex of the simplex, as is the case for some standard priors such \cite{Pose21}.  

\subsection{Stationarity Analysis of a Semiconductor Laser System with Optical Feedback}

One of the core (often implicit) assumptions made when computing permutation entropy is that the associated probability distribution, $\mathbf{P}$, is time-invariant. The possibility of $\mathbf{P}$, and thus the PE, changing over the course of a time series is therefore often overlooked, and dynamical changes occurring within the system under study remain undetected.

Here, it will be demonstrated that the stationarity of $H$ can be efficiently tested through Bayesian estimation, by dividing the time series into partitions and computing $p(H|O)$ for each. Changes in $H$ are detected where $p(H|O)$ have negligible overlap, and quantified via standard statistical inference.
\begin{figure}
	\includegraphics[width=8.6cm]{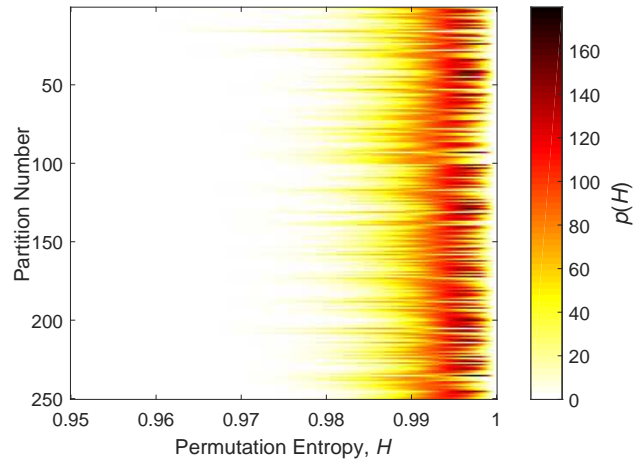}
	\caption{\label{fig:WhiteNoisePDF} Estimated $p(H)$ for a white-noise time series using hyperparameters $\alpha_i = 1$ for the Dirichlet prior distribution and $D = 3$ for 250 partitions of white-noise time series consisting of 1,000 points each.}
\end{figure}

A white-noise time series of 250,000 points was initially generated and partitioned into sections of 1,000 points each. Setting a Dirichlet prior with $\alpha_i = 1$, the posterior distribution, $p(\mathbf{P})$ was computed by adding the observed counts of $\pi_i$ ($D$ = 3) to each $\alpha_i$ (i.e. so $\alpha_i$ of the posterior distribution were set to $\pi_i + 1$). Using the procedure outlined in section \ref{sec:Reconstructing}, $p(H|O)$ was reconstructed for each partition and then plotted in Fig. \ref{fig:WhiteNoisePDF}, where the vertical axis (the partition number) can essentially be interpreted as a time coordinate, thus this can be viewed as a visualization of $p(H|O)$ varying with time. Although there is some fluctuation in the estimate of $p(H)$ between partitions as expected, substantial overlap (self-similarity) between these distributions remain, thus it would be concluded that the visualization in Fig. \ref{fig:WhiteNoisePDF} is consistent with a stationary system; an expected result given the random process used to generate the data.

Armed with this understanding, this same analysis can be applied to power time-series collected from a semiconductor laser with optical feedback (SLWOF) system. Details of this system and the procedures used to acquire the data have been reported previously \cite{Toomey14}, and are known to produce outputs with complex, potentially chaotic, outputs. 

\begin{figure}
	\includegraphics[width=8.6cm]{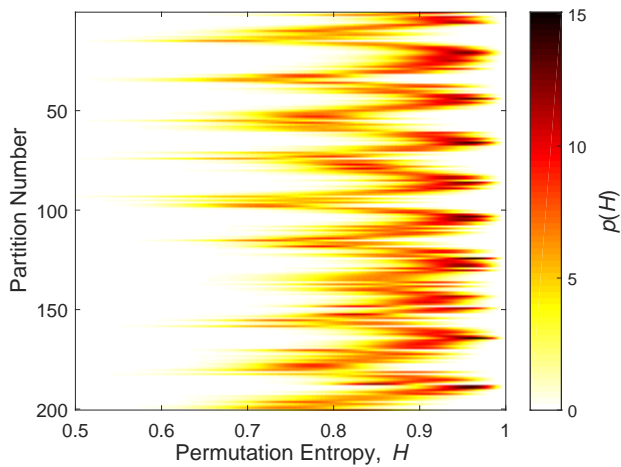}
	\caption{\label{fig:SWLOFPDF} Estimated $p(H)$ for a SLWOF time series (drive current 68.5 mA, optical feedback 8.5 \%) using hyperparameters $\alpha_i = 1$ for the Dirichlet prior distribution and $D = 3$ for 200 partitions of SWLOF power time-series consisting of 1,000 points each.}
\end{figure}

The same parameters and partitioning was used here as for the white-noise case, except the embedding delay was set to $\tau = 5$ in order to minimize short-range correlations within the data, so that the sequence of symbol observations is as independent as possible. From Fig. \ref{fig:SWLOFPDF} it can be seen that there are periodic fluctuations in $p(H|O)$, in stark contrast to the white-noise case. Because the overlap of $p(H|O)$ between the maximum and minimum portions of the cycle is negligible, we conclude that the system is non-stationary. The frequency of this oscillation in $H$ is around 1 MHz, which appears to be linked to the cavity round-trip time of the laser. The exact mechanism driving this behavior is a topic of current work.

\section{Conclusion}

\label{sec:conclusion}

A method for efficiently estimating the probability distribution of PE using a Bayesian approach has been presented for systems where the observation of dynamical states (represented by ordinal patterns) can be modeled as an iid multinomial random process. The supposition of iid randomness enables the use of Dirichlet conjugate prior distributions in $p(\mathbf{P})$ to compute the moments of the resultant posterior distribution of $p(H|O)$. It is numerically demonstrated that $p(H|O)$ are well-described by standard Beta distributions, and can thus be fully reconstructed from its moments.

With this methodology, the stationarity of $H$ was tested by computing $p(H|O)$ over a series of 1,000-point partitions taken from the original data set. This test proved effective at visualizing changes in $p(O|H)$ and thus detecting non-stationary behavior in a semiconductor-laser with optical-feedback (SWLOF) system. By contrast, white-noise was found to be consistent with a stationary system (as expected). While variations in $p(H|O)$ were observed, they were not statistically significant enough (i.e. the overlap between $p(H|O)$ remained large) to rule out the null hypothesis of stationary $H$.  

Overall, this Bayesian approach enables the uncertainty associated with permutation entropy calculations to be articulated in a rigorous framework. Even data sets consisting of only a few observations can be usefully analyzed by virtue of the incorporated prior distributions. This not just useful for determining whether dynamical state probabilities are stationarity, it can be applied to any statistical inference problem where the dynamical state observations can be modeled as an iid random process. 
 
\begin{acknowledgments}
This research was supported by the Science and Industry Endowment Fund, Australia, Research project RP 04-174, and Macquarie University.
\end{acknowledgments}

\bibliography{References}

\section{Appendix}

\subsection{Third Standardized Moment}

The third standardized moment, commonly referred to as the skewness, $\gamma$, is defined in terms of the third moment as
\begin{equation}
\gamma = \frac{1}{\sigma^3}(E[H^3] - 3\mu\sigma^2 - \mu^3),
\end{equation}
where $\mu$ is the mean and $\sigma$ is the standard deviation. From Eq. \ref{eq:Betas}, the third moment can be expressed as
\begin{equation}
\label{eq:Beta3}
E[H^3] = \frac{-1}{\log(D!)^3}\sum_{|\mathbf{k}|=3}{3\choose \mathbf{k}}\frac{1}{\mathbf{B}(\boldsymbol{\alpha})}\frac{\partial^3}{\partial\boldsymbol{\alpha}^\mathbf{k}}\mathbf{B}(\boldsymbol{\alpha}+\mathbf{k}).
\end{equation}
This summation can be split into three parts, $S_1$, where the multi-index has a single non-zero component (a 3), $S_2$, where the multi-index has two non-zero components (a 1 and a 2), and $S_3$, where the multi-index has three non-zero components (all 1's),
\begin{equation}
\label{eq:TM}
E[H^3] = -\frac{1}{\log(D!)^3}(S_1 + S_2 + S_3).
\end{equation}
The first summation, $S_1$ is given by
\begin{equation}
S_1 = \sum_{i=1}^{D!}\frac{\Gamma(\alpha_0)}{\Gamma(\alpha_i)}\frac{\partial^3}{\partial\alpha_i^3}\frac{\Gamma(\alpha_i + 3)}{\Gamma(\alpha_0 + 3)}.
\end{equation}
Evaluating the derivative yields
\begin{equation}
\begin{split}
S_1 = \sum_{i=1}^{D!}A_i\bigg(\Delta\psi(\alpha_i+3)^3 + \Delta\psi_2(\alpha_i+3)\\+ 3\Delta\psi_1(\alpha_i+3)\Delta\psi(\alpha_i+3)\bigg).
\end{split}
\end{equation}
where
\begin{equation}
A_i = \frac{\alpha_i(\alpha_i+1)(\alpha_i+2)}{\alpha_0(\alpha_0+1)(\alpha_0+2)}.
\end{equation}
and
\begin{equation}
\Delta\psi_m(\alpha_i + k_i) = \psi_m(\alpha_i + k_i) - \psi_m(\alpha_0 + n)
\end{equation}
as before. The second summation, $S_2$, is given by
\begin{equation}
\begin{split}
S_2 = \sum_{i,j=1; i \neq j}^{D!}&\frac{\Gamma(\alpha_0)}{\Gamma(\alpha_i)\Gamma(\alpha_j)} \times \\ &\frac{\partial^3}{\partial\alpha_i^2\partial\alpha_j}\frac{\Gamma(\alpha_i + 2)\Gamma(\alpha_j + 1)}{\Gamma(\alpha_0 + 3)}.
\end{split}
\end{equation}
Evaluating the derivatives, $S_2$ can be expressed as;
\begin{equation}
\begin{split}
S_2 = \sum_{i,j = 1; i \neq j}^{D!}3B_i\bigg(\Delta\psi(\alpha_i+2)^2\Delta\psi(\alpha_j+1)\\+ \Delta\psi_1(\alpha_i+2)\Delta\psi(\alpha_j+1)\\-
2\Delta\psi(\alpha_i+2)\psi_1(\alpha_0+3) - \psi_2(\alpha_0+3)\bigg),
\end{split}
\end{equation}
Noting the multinomial coefficient of 3, which persists in this instance despite summing over permutations. The coefficient $B_{ij}$ is given by
\begin{equation}
B_{ij} = \frac{\alpha_i(\alpha_i+1)\alpha_j}{\alpha_0(\alpha_0+1)(\alpha_0+2)}.
\end{equation}
Finally, the summation $S_3$, given by
\begin{equation}
\begin{split}
S_3 = &\sum_{i,j,k=1; i \neq j \neq k}^{D!}\frac{\Gamma(\alpha_0)}{\Gamma(\alpha_i)\Gamma(\alpha_j)\Gamma(\alpha_k)} \times \\ &\frac{\partial^3}{\partial\alpha_i\partial\alpha_j\partial\alpha_k}\frac{\Gamma(\alpha_i + 1)\Gamma(\alpha_j + 1)\Gamma(\alpha_k + 1)}{\Gamma(\alpha_0 + 3)},
\end{split}
\end{equation}
can be expressed as
\begin{equation}
\begin{split}
S_3 = \sum_{i,j,k=1; i \neq j \neq k}^{D!}&C_{ijk}\bigg(-\psi_2(\alpha_0 + 3)\\
- \big[\Delta\psi(\alpha_i+1) &+ \Delta\psi(\alpha_j+1) + \\
+ \Delta\psi(&\alpha_k+1)\big]\psi(\alpha_0+3) \\
+ \Delta\psi(\alpha_i+1)&\Delta\psi(\alpha_j+1)\Delta\psi(\alpha_k+1)\bigg),
\end{split}
\end{equation}
where
\begin{equation}
C_{ijk} = \frac{\alpha_i\alpha_j\alpha_k}{\alpha_0(\alpha_0+1)(\alpha_0+2)}.
\end{equation}

\subsection{Fourth Standardized Moment}

The fourth standardized moment, commonly referred to as the kurtosis, $\kappa$, is defined in terms of the fourth moment as;
\begin{equation}
\kappa = \frac{1}{\sigma^4}(E[H^4] - 4\gamma\mu\sigma^3 - 6\mu^2\sigma^2 - \mu^4).
\end{equation}
where $\mu$ is the mean, $\sigma$ is the standard deviation and $\gamma$ is the skewness. The fourth moment can be expressed using Eq. \ref{eq:Betas} as
\begin{equation}
E[H^4] = \frac{1}{\log(D!)^4}\sum_{|\mathbf{k}|=4}{4\choose \mathbf{k}}\frac{1}{\mathbf{B}(\boldsymbol{\alpha})}\frac{\partial^4}{\partial\boldsymbol{\alpha}^\mathbf{k}}\mathbf{B}(\boldsymbol{\alpha}+\mathbf{k}).
\end{equation}
As before, the summation can be split into multiple parts
\begin{equation}
\label{eq:FoM}
E[H^4] = \frac{1}{\log(D!)^4}(S_1 + S_2 + S_3 + S_4 + S_5).
\end{equation}
The summation $S_1$ consists of multi-indices with one non-zero term (a 4), $S_2$ consists of multi-indices with two equal non-zero terms (two 2s), $S_3$ consists of multi-indices with two unequal non-zero terms (a 3 and a 1), $S_4$ consists of multi-indices with three unequal non-zero terms (a 2 and two 1s) and $S_5$ consists of multi-indices with four non-zero terms (four 1s).

$S_1$ is given by
\begin{equation}
S_1 = \sum_{i=1}^{D!}\frac{\Gamma(\alpha_0)}{\Gamma(\alpha_i)}\frac{\partial^4}{\partial\alpha_i^4}\frac{\Gamma(\alpha_i + 4)}{\Gamma(\alpha_0 + 4)},
\end{equation}
which evaluates to
\begin{equation}
\begin{split}
S_1 = \sum_{i=1}^{D!}A_i\bigg(&\Delta\psi(\alpha_i + 4) + \Delta\psi_3(\alpha_i + 4)\\+ 6&\Delta\psi_1(\alpha_i + 4)\Delta\psi(\alpha_i + 4)^2\\ + &\Delta\psi_2(\alpha_i + 4)\Delta\psi(\alpha_i + 4)\\ + 3\Delta\psi_2(\alpha_i + 4)&\Delta\psi_1(\alpha_i + 4)^2\Delta\psi(\alpha_i + 4)\bigg),
\end{split}
\end{equation}
where
\begin{equation}
A_i = \frac{\alpha_i(\alpha_i+1)(\alpha_i+2)(\alpha_i+3)}{\alpha_0(\alpha_0+1)(\alpha_0+2)(\alpha_0+3)}.
\end{equation}
Next, $S_2$ is given by
\begin{equation}
\begin{split}
S_2 = \sum_{i=1,j; i \neq j}^{D!}&\frac{\Gamma(\alpha_0)}{\Gamma(\alpha_i)\Gamma(\alpha_j)} \times \\&\frac{\partial^4}{\partial\alpha_i^2\alpha_j^2}\frac{\Gamma(\alpha_i + 2)(\alpha_j + 2)}{\Gamma(\alpha_0 + 4)}.
\end{split}
\end{equation}
Evaluating the derivatives yields
\begin{equation}
\begin{split}
S_2 = \sum_{i,j=1; i \neq j}^{D!}B_{ij}\bigg([\Delta\psi_1(\alpha_i + 2) + \Delta\psi(\alpha_i + 2)^2] \times \\ [\Delta\psi_1(\alpha_j + 2) + \Delta\psi(\alpha_j + 2)^2] \\ -4\Delta\psi(\alpha_i + 2)\Delta\psi(\alpha_j + 2)\psi_1(\alpha_0 + 4) \\ -2(\Delta\psi(\alpha_i + 2) + \Delta\psi(\alpha_j + 2))\psi_2(\alpha_0 + 4) \\ +2\psi_1(\alpha_0 + 4)^2 - \psi_3(\alpha_0 + 4)\bigg),
\end{split}
\end{equation}
where
\begin{equation}
B_{ij} = \frac{\alpha_i(\alpha_i+1)\alpha_j(\alpha_j+1)}{\alpha_0(\alpha_0+1)(\alpha_0+2)(\alpha_0+3)}.
\end{equation}
The expression for $S_3$ is
\begin{equation}
\begin{split}
S_3 = \sum_{i=1,j; i \neq j}^{D!}&\frac{\Gamma(\alpha_0)}{\Gamma(\alpha_i)\Gamma(\alpha_j)} \times \\&\frac{\partial^4}{\partial\alpha_i^3\alpha_j}\frac{\Gamma(\alpha_i + 3)(\alpha_j + 1)}{\Gamma(\alpha_0 + 4)},
\end{split}
\end{equation}
which evaluates to
\begin{equation}
\begin{split}
S_3 = \sum_{i,j=1; i \neq j}^{D!}C_{ij}\bigg(\Delta\psi(\alpha_i + 3)^3\Delta\psi(\alpha_j + 1) \\ + 3\Delta\psi(\alpha_i + 3)\Delta\psi(\alpha_j + 1)\Delta\psi_1(\alpha_i + 3) \\ - \Delta\psi_1(\alpha_i + 3)\psi_1(\alpha_0 + 4) \\ - 3\Delta\psi(\alpha_i + 3)^2\psi_1(\alpha_0 + 4) \\ + \Delta\psi(\alpha_j + 1)\Delta\psi_2(\alpha_i + 3) \\ - 3\Delta\psi(\alpha_i + 3)\psi_2(\alpha_0 + 4) - \psi_3(\alpha_0 + 4)\bigg),
\end{split}
\end{equation}
where
\begin{equation}
C_{ij} = \frac{\alpha_i(\alpha_i+1)(\alpha_i+2)\alpha_j}{\alpha_0(\alpha_0+1)(\alpha_0+2)(\alpha_0+3)}.
\end{equation}
Next, the expression for $S_4$ is
\begin{equation}
\begin{split}
S_4 = &\sum_{i=1,j,k; i \neq j \neq k}^{D!}\frac{\Gamma(\alpha_0)}{\Gamma(\alpha_i)\Gamma(\alpha_j)\Gamma(\alpha_k)} \times \\&\frac{\partial^4}{\partial\alpha_i^2\alpha_j\alpha_k}\frac{\Gamma(\alpha_i + 2)(\alpha_j + 1)(\alpha_k + 1)}{\Gamma(\alpha_0 + 4)}.
\end{split}
\end{equation}
This evaluates to
\begin{equation}
\begin{split}
S_4 = \sum_{i,j,k=1; i \neq j \neq k}^{D!}D_{ijk}\bigg(2\psi_1(\alpha_0 + 4)^2 - \psi_3(\alpha_0 + 4) \\ + \Delta\psi(\alpha_j + 1)\Delta\psi(\alpha_k + 1)(\Delta\psi_1(\alpha_i + 2) + \Delta\psi(\alpha_i + 2)^2) \\ -(2\Delta\psi(\alpha_i + 2) + \Delta\psi(\alpha_j + 1) + \Delta\psi(\alpha_k + 1))\psi_2(\alpha_0 + 4) \\ -(\Delta\psi_1(\alpha_i + 2) + \Delta\psi(\alpha_i + 2)^2)\psi_1(\alpha_0 + 4) \\ - 2\Delta\psi(\alpha_i + 2)\Delta\psi(\alpha_j + 1)\psi_1(\alpha_0 + 4) \\ - 2\Delta\psi(\alpha_i + 2)\Delta\psi(\alpha_k + 1)\psi_1(\alpha_0 + 4)\bigg),
\end{split}
\end{equation}
where
\begin{equation}
D_{ijk} = \frac{\alpha_i(\alpha_i+1)\alpha_j\alpha_k}{\alpha_0(\alpha_0+1)(\alpha_0+2)(\alpha_0+3)}.
\end{equation}
Finally, $S_5$ is given by
\begin{equation}
\begin{split}
S_5 = &\sum_{i=1,j,k,l; i \neq j \neq k \neq l}^{D!}\frac{\Gamma(\alpha_0)}{\Gamma(\alpha_i)\Gamma(\alpha_j)\Gamma(\alpha_k)\Gamma(\alpha_l)} \times \\&\frac{\partial^4}{\partial\alpha_i\alpha_j\alpha_k\alpha_l}\frac{\Gamma(\alpha_i + 1)(\alpha_j + 1)(\alpha_k + 1)(\alpha_l + 1)}{\Gamma(\alpha_0 + 4)}.
\end{split}
\end{equation}
This evaluates to
\begin{equation}
\begin{split}
S_5 = \sum_{i,j,k,l=1; i \neq j \neq k \neq l}^{D!}E_{ijkl}\bigg(\Delta\psi(\alpha_i + 1)\Delta\psi(\alpha_j + 1) \times \\ \Delta\psi(\alpha_k + 1)\Delta\psi(\alpha_l + 1) \\ -\Delta\psi(\alpha_i + 1)\Delta\psi(\alpha_j + 1)\psi_1(\alpha_0 + 4) \\ -\Delta\psi(\alpha_i + 1)\Delta\psi(\alpha_k + 1)\psi_1(\alpha_0 + 4) \\ -\Delta\psi(\alpha_i + 1)\Delta\psi(\alpha_l + 1)\psi_1(\alpha_0 + 4) \\ -\Delta\psi(\alpha_j + 1)\Delta\psi(\alpha_k + 1)\psi_1(\alpha_0 + 4) \\ -\Delta\psi(\alpha_j + 1)\Delta\psi(\alpha_l + 1)\psi_1(\alpha_0 + 4) \\ -\Delta\psi(\alpha_k + 1)\Delta\psi(\alpha_l + 1)\psi_1(\alpha_0 + 4) \\ -[\Delta\psi(\alpha_i + 1) + \Delta\psi(\alpha_j + 1) \\ + \Delta\psi(\alpha_k + 1) + \Delta\psi(\alpha_l + 1)]\psi_2(\alpha_0 + 4) \\ + 3\psi_1(\alpha_0 + 4)^2 - \psi_3(\alpha_0 + 4)\bigg),
\end{split}
\end{equation}
where
\begin{equation}
E_{ijkl} = \frac{\alpha_i\alpha_j\alpha_k\alpha_l}{\alpha_0(\alpha_0+1)(\alpha_0+2)(\alpha_0+3)}.
\end{equation}

\end{document}